\input harvmac

%
\let\includefigures=\iftrue
%
%
%
\newfam\black
\input rotate
\input epsf
\noblackbox
%
%
\includefigures
\message{If you do not have epsf.tex (to include figures),}
\message{change the option at the top of the tex file.}
\def\figin{\epsfcheck\figin}\def\figins{\epsfcheck\figins}
\def\epsfcheck{\ifx\epsfbox\UnDeFiNeD
\message{(NO epsf.tex, FIGURES WILL BE IGNORED)}
\gdef\figin##1{\vskip2in}\gdef\figins##1{\hskip.5in}
\else\message{(FIGURES WILL BE INCLUDED)}%
\gdef\figin##1{##1}\gdef\figins##1{##1}\fi}
\def\DefWarn#1{}
\def\N{{\cal N}}
\def\figinsert{\goodbreak\midinsert}
\def\ifig#1#2#3{\DefWarn#1\xdef#1{fig.~\the\figno}
\writedef{#1\leftbracket fig.\noexpand~\the\figno}%
\figinsert\figin{\centerline{#3}}\medskip\centerline{\vbox{\baselineskip12pt
\advance\hsize by -1truein\noindent\footnotefont{\bf
Fig.~\the\figno:} #2}}
\bigskip\endinsert\global\advance\figno by1}
\else
\def\ifig#1#2#3{\xdef#1{fig.~\the\figno}
\writedef{#1\leftbracket fig.\noexpand~\the\figno}%
\global\advance\figno by1} \fi
\def\yboxit#1#2{\vbox{\hrule height #1 \hbox{\vrule width #1
\vbox{#2}\vrule width #1 }\hrule height #1 }}
\def\fillbox#1{\hbox to #1{\vbox to #1{\vfil}\hfil}}
\def\ybox{{\lower 1.3pt \yboxit{0.4pt}{\fillbox{8pt}}\hskip-0.2pt}}

\def\rightarrowbox#1#2{
  \setbox1=\hbox{\kern#1{${ #2}$}\kern#1}
  \,\vbox{\offinterlineskip\hbox to\wd1{\hfil\copy1\hfil}
    \kern 3pt\hbox to\wd1{\rightarrowfill}}}

\def\half{{1\over 2}}

\def\tr{{\rm tr\ }}

\def\CL{{\cal L}}

\def\CO{{\cal O}}

\def\tilde{\widetilde}

\def\II{\relax{I\kern-.10em I}}

\def\bar{\overline}

\def\IZ{\relax\ifmmode\mathchoice
{\hbox{\cmss Z\kern-.4em Z}}{\hbox{\cmss Z\kern-.4em Z}}
{\lower.9pt\hbox{\cmsss Z\kern-.4em Z}} {\lower1.2pt\hbox{\cmsss
Z\kern-.4em Z}}\else{\cmss Z\kern-.4em Z}\fi}
\def\IB{\relax{\rm I\kern-.18em B}}
\def\IC{{\relax\hbox{$\inbar\kern-.3em{\rm C}$}}}
\def\ID{\relax{\rm I\kern-.18em D}}
\def\IE{\relax{\rm I\kern-.18em E}}
\def\IF{\relax{\rm I\kern-.18em F}}
\def\IG{\relax\hbox{$\inbar\kern-.3em{\rm G}$}}
\def\IGa{\relax\hbox{${\rm I}\kern-.18em\Gamma$}}
\def\IH{\relax{\rm I\kern-.18em H}}
\def\II{\relax{\rm I\kern-.18em I}}
\def\IK{\relax{\rm I\kern-.18em K}}
\def\IN{\relax{\rm I\kern-.18em N}}
\def\IP{\relax{\rm I\kern-.18em P}}

%
\def\inbar{\,\vrule height1.5ex width.4pt depth0pt}

\font\cmss=cmss10 \font\cmsss=cmss10 at 7pt
\def\IR{\relax{\rm I\kern-.18em R}}

\def\lp10{l_P^{10}}
\def\lp11{l_P^{11}}
\def\R11{R_{11}}

\def\alphadot{{\dot \alpha}}
\def\betadot{{\dot \beta}}
\def\gammadot{{\dot \gamma}}

\def\thetabar{{\bar\theta}}

\def\tilde{\widetilde}

%
%
\lref\WessCP{ J.~Wess and J.~Bagger, ``Supersymmetry And
Supergravity,'' Princeton, USA: Univ. Pr. (1992).
}

\lref\SchwarzPF{ J.~H.~Schwarz and P.~Van Nieuwenhuizen,
``Speculations Concerning A Fermionic Substructure Of
Space-Time,'' Lett.\ Nuovo Cim.\  {\bf 34}, 21 (1982).
}

\lref\BerkovitsBF{ N.~Berkovits, ``A new description of the
superstring,'' arXiv:hep-th/9604123.
}

\lref\BrinkNZ{ L.~Brink and J.~H.~Schwarz, ``Clifford Algebra
Superspace,'' CALT-68-813
}

\lref\SeibergVS{ N.~Seiberg and E.~Witten, ``String theory and
noncommutative geometry,'' JHEP {\bf 9909}, 032 (1999)
[arXiv:hep-th/9908142].
}

\lref\KlemmYU{ D.~Klemm, S.~Penati and L.~Tamassia,
``Non(anti)commutative superspace,'' arXiv:hep-th/0104190.
}

\lref\ChepelevGA{ I.~Chepelev and C.~Ciocarlie, ``A path integral
approach to noncommutative superspace,'' arXiv:hep-th/0304118.
}

\lref\deBoerDN{ J.~de Boer, P.~A.~Grassi and P.~van Nieuwenhuizen,
``Non-commutative superspace from string theory,''
arXiv:hep-th/0302078.
}

\lref\KawaiYF{ H.~Kawai, T.~Kuroki and T.~Morita, ``Dijkgraaf-Vafa
theory as large-N reduction,'' arXiv:hep-th/0303210.
}

\lref\CasalbuoniHX{ R.~Casalbuoni, ``Relativity And
Supersymmetries,'' Phys.\ Lett.\ B {\bf 62}, 49 (1976).
}

\lref\CasalbuoniBJ{ R.~Casalbuoni, ``On The Quantization Of
Systems With Anticommutating Variables,'' Nuovo Cim.\ A {\bf 33},
115 (1976).
}

\lref\CasalbuoniTZ{ R.~Casalbuoni, ``The Classical Mechanics For
Bose-Fermi Systems,'' Nuovo Cim.\ A {\bf 33}, 389 (1976).
}

\lref\OoguriQP{ H.~Ooguri and C.~Vafa, ``The C-deformation of
gluino and non-planar diagrams,'' arXiv:hep-th/0302109.
}

\lref\OoguriTT{ H.~Ooguri and C.~Vafa, ``Gravity induced
C-deformation,'' arXiv:hep-th/0303063.
}

\lref\AbbaspurXJ{ R.~Abbaspur, ``Generalized noncommutative
supersymmetry from a new gauge symmetry,'' arXiv:hep-th/0206170.
}

\lref\FerraraMM{ S.~Ferrara and M.~A.~Lledo, ``Some aspects of
deformations of supersymmetric field theories,'' JHEP {\bf 0005},
008 (2000) [arXiv:hep-th/0002084].
}

\lref\DouglasBA{ M.~R.~Douglas and N.~A.~Nekrasov,
``Noncommutative field theory,'' Rev.\ Mod.\ Phys.\  {\bf 73}, 977
(2001) [arXiv:hep-th/0106048].
}

\lref\DavidKE{ J.~R.~David, E.~Gava and K.~S.~Narain, ``Konishi
anomaly approach to gravitational F-terms,'' arXiv:hep-th/0304227.
}

\noblackbox

\newbox\tmpbox\setbox\tmpbox\hbox{\abstractfont
}
 \Title{\vbox{\baselineskip12pt\hbox to\wd\tmpbox{\hss
 hep-th/0305248} }}
 {\vbox{\centerline{Noncommutative Superspace, $\N={1\over 2}$ Supersymmetry, }
 \centerline{Field Theory and String Theory}}
 }
\smallskip
\centerline{Nathan Seiberg}
\smallskip
\bigskip
\centerline{School of Natural Sciences, Institute for Advanced
Study, Princeton NJ 08540 USA}
\bigskip
\vskip 1cm
 \noindent
We deform the standard four dimensional $\N=1$ superspace by
making the odd coordinates $\theta$ not anticommuting, but
satisfying a Clifford algebra.  Consistency determines the other
commutation relations of the coordinates.  In particular, the
ordinary spacetime coordinates $x$ cannot commute.  We study
chiral superfields and vector superfields and their interactions.
As in ordinary noncommutative field theory, a change of variables
allows us to express the gauge interactions in terms of component
fields which are subject to standard gauge transformation laws.
Unlike ordinary noncommutative field theories, the change of the
Lagrangian is a polynomial in the deformation parameter.  Despite
the deformation, the noncommutative theories still have an
antichiral ring with all its usual properties.  We show how these
theories with precisely this deformation arise in string theory in
a graviphoton background.

\Date{May 2003}
%

%
%

\newsec{Introduction}

One of the objectives of string theory is to understand the nature
of spacetime.  It is therefore natural to ask how to add structure
to standard $\IR^4$, and how this space can be deformed.  An
example of an added structure is the addition of anticommuting
spinor coordinates $\theta^\alpha$ and $\thetabar^\alphadot$ (we
follow the notation of \WessCP). The resulting space is known as
superspace.  An example of a deformation is to make $\IR^4$
noncommutative (for a review, see e.g.\ \DouglasBA). The
combination of these two ideas; i.e.\ that the anticommuting
coordinates $\theta$ form a Clifford algebra
 \eqn\ththa{\{\theta^\alpha,\theta^\beta\} =C^{\alpha \beta}}
has been explored by many people (for a partial list, see
\refs{\CasalbuoniHX\CasalbuoniBJ\CasalbuoniTZ
\BrinkNZ\SchwarzPF\FerraraMM\KlemmYU\AbbaspurXJ \deBoerDN\OoguriQP
\OoguriTT\KawaiYF\ChepelevGA-\DavidKE}).   Here we will examine
the consequences of this deformation.  We will see that because of
\ththa\ the $\IR^4$ coordinates $x^\mu$ cannot commute. Instead,
the chiral coordinates
 \eqn\ydefa{y^\mu=x^\mu+i\theta^\alpha
 \sigma^\mu_{\alpha\alphadot} \thetabar^\alphadot}
can be taken to commute.  Our noncommutative superspace is defined
as follows. All the (anti)commutators of $y^\mu$, $\theta^\alpha$
and $\thetabar^\alphadot$ vanish except \ththa.  We will refer to
the space with nonzero $C$ as noncommutative and with zero $C$ as
commutative (although in fact, it is partially commutative and
partially anticommutative).

In section 2 we will motivate the definition of the noncommutative
superspace and will explore some of its properties.  In
particular, we will show that the deformation \ththa\ breaks half
the supersymmetry. Only the $Q_\alpha$ supercharges are conserved,
while the $\bar Q_\alphadot$ supercharges are broken. Since half
of $\N=1$ supersymmetry is broken, we can refer to the unbroken
$Q$ supersymmetry as $\N={1\over 2}$ supersymmetry.  We will also
mention some generalizations of the minimal deformation. In
section 3 we will study chiral and antichiral superfields and
their interactions. Even though the deformation \ththa\ breaks
Lorentz invariance, $\int d^2\theta W(\Phi)$ turns out to be
invariant.  $\int d^2\thetabar \bar W(\bar \Phi)$ turns out to be
independent of $C$.

In section 4 we study vector superfields. Because of the
deformation \ththa, the standard gauge transformation rules are
deformed.  However, a simple change of variables allows us to use
standard component fields with standard gauge transformation
rules.  In the Wess-Zumino gauge we find that the standard
Lagrangian $i\tau \int d^2\theta ~\tr~ W W -i\bar\tau \int
d^2\thetabar ~\tr~ \bar W\bar W$ is deformed by $(i\tau -i \bar
\tau)\left( -i C^{\mu\nu}\tr ~ F_{\mu\nu} \bar\lambda\bar\lambda +
{|C|^2 \over 4} \tr ~ (\bar\lambda\bar\lambda)^2\right)+ {\rm
total ~ derivative}$, where $C^{\mu\nu}\equiv C^{\alpha\beta}
\epsilon_{\beta\gamma} \sigma^{\mu\nu ~\gamma}_\alpha$.

Section 5 is devoted to the study of the various rings. The chiral
ring of the commutative theory is ruined by the deformation.
However, the antichiral ring is still present even though we have
only half of the supersymmetry. The antichiral ring retains all of
the standard properties which it has in the commutative theory. In
section 6 we study instantons and anti-instantons in these
theories. Finally, in section 7 we show how these theories arise
on branes in the presence of a background graviphoton.

\newsec{Superspace}

We want to consider the consequences of the nontrivial
anticommutator
 \eqn\thth{\{\theta^\alpha,\theta^\beta\} =C^{\alpha \beta}}
on superspace. Because of \thth, functions of $\theta$ should be
ordered.  We use Weyl ordering.  This means that the function
$\theta^\alpha \theta^\beta$ in the commutative space is ordered
as $\half [\theta^\alpha, \theta^\beta]=-\half
\epsilon^{\alpha\beta}\theta\theta$ (we follow the conventions of
Wess and Bagger \WessCP) in the noncommutative space. When
functions of $\theta$ are multiplied, the result should be
reordered.  As in ordinary noncommutative geometry, this is
implemented by the star product
 \eqn\stard{\eqalign{
 f(\theta)*g(\theta) &= f(\theta)\exp\left(-
 {C^{\alpha\beta}\over 2}
 \overleftarrow{\partial\over\partial\theta^\alpha}
 \overrightarrow{\partial\over\partial\theta^\beta}\right)g(\theta) \cr
 &= f(\theta)\left(1- {C^{\alpha\beta}\over 2}
 \overleftarrow{\partial\over\partial\theta^\alpha}
 \overrightarrow{\partial\over\partial\theta^\beta} -\det C
 \overleftarrow{\partial\over\partial\theta\theta}
 \overrightarrow{\partial\over\partial\theta\theta}
 \right)g(\theta)\cr
 \overrightarrow{\partial\over
\partial\theta^\beta} \theta^\alpha & \equiv {\partial\over
\partial\theta^\beta} \theta^\alpha = \delta_\beta^\alpha \cr
\theta^\alpha \overleftarrow{\partial\over
\partial\theta^\beta} &\equiv -  \delta_\beta^\alpha \cr
 {\partial\over\partial\theta\theta}&\equiv {1\over 4}
\epsilon^{\alpha\beta} {\partial \over \partial\theta^\alpha}
{\partial \over \partial\theta^\beta}
 }}
If $f(\theta)$ is a bosonic function, $f(\theta)\overleftarrow{
\partial\over \partial\theta^\beta}=\overrightarrow{\partial\over
\partial \theta^\beta} f(\theta) $.  Since $\overleftarrow
{\partial\over\partial\theta\theta}$  and
$\overrightarrow{\partial\over\partial \theta\theta}$ are bosonic,
their definitions are obvious.

Useful examples are
 \eqn\useex{\eqalign{
 &\theta^\alpha *\theta^\beta = -{1\over 2} \epsilon^{\alpha
 \beta} \theta \theta + {1\over 2} C^{\alpha \beta} \cr
 &\theta^\alpha * \theta \theta =C^{\alpha \beta} \theta_\beta \cr
 &\theta \theta *\theta^\alpha  =-C^{\alpha \beta} \theta_\beta \cr
 &\theta \theta * \theta \theta = - {1 \over 2}
 \epsilon_{\alpha\beta} \epsilon_{\alpha'\beta'}C^{\alpha\alpha'}
 C^{\beta \beta'} = -\det C
 }}

If $f$ or $g$ depend on additional variables which have nontrivial
commutation relations, the expression \stard\ should be modified
appropriately.

Regarding the other coordinates, consider first simplest
possibility that $\thetabar^\alphadot$ satisfies standard
commutation relations
 \eqn\thbthb{\{\thetabar^\alphadot,\thetabar^\betadot\} =
 \{\thetabar^\alphadot,\theta^\beta\} =
 [ \thetabar^\alphadot,x^\mu]=0.}
This means that $\thetabar$ is not the complex conjugate of
$\theta$, which is possible only in Euclidean space. We will be
working on Euclidean $\IR^4$, but we will continue to use the
Lorentzian signature notation. Since $\thetabar$ anticommutes with
$\theta$, we find the following useful identity
 \eqn\anous{\theta \sigma^\mu \thetabar * \theta \sigma^\nu
 \thetabar = -{1 \over 2} \eta^{\mu\nu} \theta\theta\thetabar
 \thetabar -{1\over 2} \thetabar\thetabar C^{\mu\nu}}
where
 \eqn\cmunu{C^{\mu\nu}\equiv C^{\alpha\beta}
 \epsilon_{\beta\gamma} \sigma^{\mu\nu ~\gamma}_\alpha}
is selfdual.  Useful identities are
 \eqn\cmunuid{\eqalign{
 &C^{\alpha\beta}={1\over 2}\epsilon^{\alpha\gamma}
 \sigma^{\mu\nu ~\beta}_\gamma C_{\mu\nu}\cr
 &|C|^2 \equiv C^{\mu\nu} C_{\mu\nu} =4\det C}.}

What about the commutation relations of $x$?  The simplest
possibility is $[x^\mu,x^\nu]=[x^\mu, \theta^\alpha]=0$.  This
makes it difficult to define chiral and antichiral fields because
the ordinary $D$ and $\bar D$ do not act as derivations (see
below). Instead, using
 \eqn\ydef{y^\mu=x^\mu+i\theta^\alpha
 \sigma^\mu_{\alpha\alphadot} \thetabar^\alphadot}
we can accompany \thth\thbthb\ with
 \eqn\ycom{[y^\mu,y^\nu]=[y^\mu,\theta^\alpha]=[y^\mu,
 \thetabar^\alphadot]=0.}
This means that
 \eqn\xcom{\eqalign{
 &[x^\mu,\theta^\alpha]=i
 C^{\alpha\beta}\sigma^\mu_{\beta\alphadot}
 \thetabar ^\alphadot \cr
 &[x^\mu,x^\nu]= \thetabar \thetabar C^{\mu\nu}\cr}}

When functions on superspace are expressed in terms of $y,
\theta,\thetabar$ we can use the star product \stard\ where the
derivatives with respect to $\theta$ are at fixed $y$ and
$\thetabar$.  We will follow this convention and will take
$\partial\over\partial \theta$ to mean a derivative at fixed $y$
and $\thetabar$.

We can now define the covariant derivatives.  Since our $\theta$
and $\thetabar$ derivatives are at fixed $y$ (rather than at fixed
$x$), the standard expressions
 \eqn\DDbar{\eqalign{
 &D_\alpha={\partial \over \partial \theta^\alpha}
 +2i\sigma^\mu_{\alpha \alphadot}\thetabar^\alphadot
 {\partial\over \partial y^\mu} \cr
 &\bar D_\alphadot =- {\partial \over \partial
 \thetabar^\alphadot}
 }}
can still be used. Clearly, they satisfy
 \eqn\Drelations{\eqalign{
 &\{D_\alpha,D_\beta\}=0\cr
 &\{\bar D_\alphadot,\bar D_\betadot\}=0\cr
 &\{\bar D_\alphadot, D_\alpha\}= -2i\sigma^\mu_{\alpha
 \alphadot}{\partial\over \partial y^\mu}
 }}
exactly as in the commutative space with $C=0$.

The supercharges of the commutative theories are
 \eqn\QQbar{\eqalign{
 &Q_\alpha={\partial \over \partial \theta^\alpha}\cr
 &\bar Q_\alphadot =- {\partial \over \partial
 \thetabar^\alphadot}+2i\theta^\alpha\sigma^\mu_{\alpha \alphadot}
 {\partial\over \partial y^\mu},
 }}
where again the derivatives with respect to $\theta$ and
$\thetabar$ are taken at fixed $y$.  They satisfy
 \eqn\QDrelations{\eqalign{
 &\{D_\alpha,Q_\beta\}=\{\bar D_\alphadot,Q_\beta\}=
 \{D_\alpha,\bar Q_\betadot\}=
 \{\bar D_\alphadot,\bar Q_\betadot\}=0\cr
  &\{ Q_\alpha, Q_\beta\}=0\cr
 &\{\bar Q_\alphadot, Q_\alpha\}= 2i\sigma^\mu_{\alpha
 \alphadot} {\partial\over \partial y^\mu}\cr
 &\{\bar Q_\alphadot,\bar Q_\betadot\}=-4C^{\alpha\beta}
 \sigma^\mu_{\alpha\alphadot}\sigma^\nu_{\beta\betadot}
 {\partial^2\over \partial y^\mu \partial y^\nu}
 }}
All these but the last one are as in the commutative space
($C=0$). The star product \stard\ is invariant under $Q$ and
therefore we expect it to be a symmetry of the space. However,
since $\bar Q$ depends explicitly on $\theta$, it is clear that
the star product is not invariant under $\bar Q$. Therefore, $\bar
Q$ is not a symmetry of the noncommutative space\foot{We could
have attempted to add to $\bar Q_\alphadot$ terms of the form
$C^{\alpha\beta}\sigma^\mu_{\alpha\alphadot}
\sigma^\nu_{\beta\betadot} \thetabar^\betadot{\partial^2\over
\partial y^\mu \partial y^\nu}$ and $C^{\alpha\beta}
\sigma^\mu_{\alpha\alphadot} {\partial^2\over \partial y^\mu
\partial \theta^\beta}$ to remove the last term in \QDrelations.
Of particular interest is adding $-iC^{\alpha\beta}
\sigma^\mu_{\alpha\alphadot} {\partial^2\over \partial y^\mu
\partial \theta^\beta}$ which makes all the commutation relations
\QDrelations, as in the commutative space.  The problem with such
modifications of $\bar Q$ is that they include second derivatives,
and therefore they do not act on products of fields as
derivations.}.  Since half of $\N=1$ supersymmetry is broken, we
can refer to the unbroken $Q$ supersymmetry as $\N={1\over 2}$
supersymmetry.

Reasonable conditions to impose on various generalizations of this
superspace is that $D$ and $\bar D$ of \DDbar\ act as derivations
on functions of superspace and that they continue to satisfy
\Drelations. These conditions clearly forbid a deformation of the
commutators of $\thetabar$ \thbthb, because $\thetabar$ appears
explicitly in $D$. However, there is still freedom in deforming
\ycom\ to
  \eqn\ycom{[y^\mu,y^\nu]=i \Theta^{\mu\nu}, \qquad\qquad
  [y^\mu,\theta^\alpha]= \Psi^{\mu  \alpha},}
with $\Theta^{\mu\nu}$ and $\Psi^{\mu  \alpha}$ commuting and
anticommuting $c$-numbers independent of $y$, $\theta$ and
$\thetabar$. $\Theta^{\mu\nu}$ has the effect of standard
noncommutativity \DouglasBA.  Here we will not explore these
deformations and will set them to zero.

\newsec{Chiral Superfields}

Chiral superfields are defined to satisfy $\bar D_\alphadot
\Phi=0$.  This means that $\Phi$ is independent of $\thetabar$. In
components it is
 \eqn\chis{\Phi(y,\theta) = A(y) + \sqrt{2} \theta \psi(y) +
 \theta\theta F(y)}
Since
 \eqn\thets{\theta \theta =\theta^\alpha \theta_\alpha=
 -\theta^1\theta^2 + \theta^2\theta^1,}
this expression is Weyl ordered.  Two chiral superfields
$\Phi_1(y,\theta)$ and $\Phi_2(y,\theta)$ are multiplied using the
star product \stard. Clearly, the result is a function of $y$ and
$\theta$ and therefore it is a chiral superfield
 \eqn\chichim{\eqalign{
 \Phi_1(y,\theta) * \Phi_2(y,\theta)=& \Phi_1(y,\theta)
  \Phi_2(y,\theta) - C^{\alpha\beta} \psi_{\alpha 1}(y)
  \psi_{\beta 2}(y) \cr
  &\qquad +\sqrt {2}C^{\alpha\beta}  \theta_\beta (\psi_{\alpha 1}(y)
  F_2(y)  -\psi_{\alpha 2} (y) F_1(y)) \cr
  &\qquad - \det C F_1(y)F_2(y).}}

Antichiral superfields are defined to satisfy $ D_\alpha \bar
\Phi=0$.  This means that $\bar \Phi$ depends only on $\thetabar$
and
 \eqn\baryd{\bar y^\mu= y^\mu -2i\theta^\alpha
 \sigma^\mu_{\alpha\alphadot}\thetabar^\alphadot.}
Since
 \eqn\ybarc{[\bar y^\mu,\bar y^\nu]=4\thetabar\thetabar
 C^{\mu\nu},}
antichiral superfields $\bar \Phi(\bar y, \thetabar)$ have to be
ordered.  One possibility is to express them in terms of $y$ and
$\theta$ and to Weyl order the $\theta$s
 \eqn\chib{\eqalign{
 \bar \Phi(y-2i\theta \sigma\thetabar,\thetabar) =& \bar
 A(y-2i\theta \sigma\thetabar) + \sqrt{2} \thetabar
 \bar\psi(y-2i\theta \sigma\thetabar) + \thetabar\thetabar \bar
 F(y)\cr
 =&\bar A(y)+\sqrt{2} \thetabar \bar\psi(y) -2i\theta
 \sigma^\mu\thetabar \partial_\mu \bar A(y) \cr
 &+ \thetabar \thetabar \left( \bar F(y) +i\sqrt{2}\theta
 \sigma^\mu \partial_\mu \bar \psi(y)+\theta\theta
 \partial^2 \bar A(y)\right).
 }}
Alternatively, we can order the $\bar y$s and use the fact that
the antichiral superfields do not have explicit $\theta$
dependence to multiply them as
 \eqn\anmul{\eqalign{
 \bar \Phi_1(\bar y,\thetabar)*\bar \Phi_2(\bar
 y,\thetabar)= &\bar \Phi_1(\bar y,\thetabar)\exp\left( 2
 \thetabar\thetabar C^{\mu\nu} \overleftarrow{\partial\over
 \partial\bar y^\mu} \overrightarrow{\partial\over\partial \bar
 y ^\nu}\right)\bar \Phi_2(\bar y,\thetabar)\cr
 =& \bar \Phi_1(\bar y,\thetabar)\bar \Phi_2(\bar y,\thetabar) + 2
 \thetabar\thetabar C^{\mu\nu}{\partial\over\partial \bar
 y ^\mu} \bar \Phi_1(\bar y,\thetabar) {\partial\over
 \partial\bar y^\mu} \bar \Phi_2(\bar y,\thetabar)
 }}
where $\partial \over \partial \bar y$ is taken at fixed
$\thetabar$, but unlike our conventions above, it is not at fixed
$y$.  Clearly, the result is an antichiral superfield.

Using these superfields and star product multiplication we can
write a Wess-Zumino Lagrangian. The simplest one is\foot{We
normalize $\int d^2\theta ~\theta\theta= \int d^2\thetabar ~
\thetabar\thetabar=\int d^4\theta ~ \theta\theta
\thetabar\thetabar =1$.}
 \eqn\wzl{\eqalign{
 \CL=&\int d^4\theta ~\bar \Phi \Phi + \int d^2 \theta~
 ({1\over 2} m \Phi *\Phi + {1\over 3}g \Phi*\Phi*\Phi)+\int d^2
 \thetabar~ ({1\over 2} \bar m \bar\Phi * \bar\Phi +
 {1\over 3}\bar g \bar\Phi*\bar\Phi *\bar\Phi  )\cr
 =& \CL(C=0) -{1\over 3} g \det C F^3 +
 {\rm total ~derivative.}
 }}
The only correction due to the noncommutativity arises from $\int
d^2\theta ~ \Phi *\Phi *\Phi$. It is amusing that even though our
deformed superspace is not Lorentz invariant, the Lagrangian \wzl\
is Lorentz invariant.

$Q$ and $\bar Q$ of \QQbar\ act on the fields \chis\chib\ exactly
as in the commutative theory with $C=0$.  Therefore, if we want to
examine the symmetries of the noncommutative theory which is based
on the Lagrangian \wzl, it is enough to examine the symmetries of
the new term $\det C F^3 $. Clearly, $Q$ is still a symmetry, but
$\bar Q$ is broken by the noncommutativity (nonzero $C$).  This is
in accord with the general expectation based on the symmetries of
the star product.

More generally, there can be several chiral fields $\Phi_i$ and
several antichiral fields $\bar\Phi_i$ which couple through $\int
d^2\theta~ W(\Phi_i) +\int d^2\thetabar~ \bar W(\bar\Phi_i)$.  The
functions $W$ and $\bar W$ are polynomials (or can be locally
holomorphic and antiholomorphic, respectively).  Consider for
example the monomials $W=\prod_i\Phi_{n_i}$, $\bar W =\prod_i \bar
\Phi_{n_i}$.  In the commutative theory the superfields commute
and the order of multiplication is not important.  This is no
longer the case in the noncommutative theory: $\Phi_1*\Phi_2 \not=
\Phi_2*\Phi_1$, $\bar\Phi_1 *\bar\Phi_2 \not= \bar \Phi_2 *\bar
\Phi_1$. Therefore, there are several different noncommutative
theories which correspond to the same commutative theory. They are
parametrized by the coefficients of monomials which are not
totally symmetrized.  This is analogous to the standard operator
ordering ambiguity in quantizing a classical theory.  The same
classical theory can lead to different quantum theories whose
Hamiltonians differ by terms of order $\hbar$.

It is natural to limit the discussion to noncommutative theories
in which these new parameters vanish.  Equivalently, we symmetrize
(Weyl order) all the products of superfields in the
superpotential.

Let us consider first $\int d^2\thetabar~ \bar W(\bar\Phi_i)$.
Without loss of generality we can focus on a monomial $\bar W
=\prod_i^* \bar \Phi_{n_i}$ where $\prod^*$ stands for a
symmetrized star product.  It is clear from \anmul\ that every
power of $C$ comes with $\thetabar\thetabar$. Therefore, there can
be at most one factor of $C$.  It is also clear from \anmul\ that
the term with $C$ is antisymmetric under the exchange of the two
superfields, and therefore it vanishes in the symmetrized product.
We conclude that $\int d^2\thetabar~ \bar W(\bar\Phi_i)$ is not
deformed in the noncommutative theory.

$\int d^2\theta~ W(\Phi_i)$ is a bit more interesting.  Consider
again a monomial $ W =\prod_i^*  \Phi_{n_i}$.  Using \chichim\ we
can expand it in powers of $C$.  Each $\Phi_n$ has at most two
${\partial \over \partial \theta^\alpha}$ derivatives acting on
it. Each factor with two derivatives ${\partial ^2 \over \partial
\theta^\alpha \partial\theta^\beta} \Phi_n= -2
\epsilon_{\alpha\beta} {\partial \over \partial \theta\theta}
\Phi_n$ is a Lorentz singlet.  The factors with one derivative are
of the form ${\partial \over\partial \theta^\alpha} \Phi_n$. Since
the index $n$ is symmetrized, the index $\alpha$ must be
antisymmetrized. Therefore, there are at most two such factors and
they are coupled to a Lorentz singlet.  Since all the derivatives
are coupled to a Lorentz singlet, the same should be true for the
product of $C$'s. We conclude that there is an even number of
$C$'s and they are coupled to a Lorentz singlet; i.e.\ the
deformation depends only on $\det C$. In other words, even though
$C$ breaks Lorentz invariance, the theory with an arbitrary
superpotential is Lorentz invariant.

\newsec{Vector Superfields}

Vector superfields $V$ describe gauge fields.  In general $V$ is a
matrix and the gauge symmetry acts as\foot{As we said above, we
follow the conventions of Wess and Bagger \WessCP.  More standard
conventions for gauge fields are obtained with the substitution
$V\to -2V$.}
 \eqn\gaua{e^V \to e^{V'} = e^{-i\bar \Lambda} e^V e^{i\Lambda},}
or infinitesimally
 \eqn\gauai{\delta e^V = -i\bar \Lambda e^V + ie^V \Lambda ,}
where $\Lambda $ and $\bar \Lambda$ are matrices of chiral and
antichiral superfields respectively.  In our case we still use
\gaua\gauai\ but the multiplication is the star product ($e^V$
means a power series in $V$, where each term is multiplied using
the star product)\foot{In order not to clutter the equations we
will suppress the star symbol.  It will be clear which product is
a star product.}.  It is important for the consistency of the
theory that our star product is such that a function of
(anti)chiral superfields is (anti)chiral. We can use the standard
expressions for the chiral and antichiral field strength
superfields
 \eqn\WWalpha{\eqalign{
 &W_\alpha =-{1\over 4} \bar D\bar D e^{-V} D_\alpha e^V \cr
 &\bar W_\alphadot ={1\over 4} D D e^V \bar D_\alphadot e^{-V} ,}}
but with star products.  As for $C=0$ they transform under \gaua\
as
 \eqn\WWalphag{\eqalign{
 &W_\alpha \to e^{- i\Lambda} W_\alpha e^{i\Lambda} \cr
 & \bar W_\alphadot \to e^{- i\bar\Lambda} \bar W_\alphadot
 e^{i\bar\Lambda},}}
or infinitesimally
 \eqn\WWalphagi{\eqalign{
 &\delta W_\alpha =- i[\Lambda , W_\alpha ] \cr
 & \delta \bar W_\alphadot = - i[ \bar\Lambda , \bar W_\alphadot].
 }}

The gauge freedom can be used to set some of the components of $V$
to zero.  For $C=0$ a convenient choice is the Wess-Zumino gauge.
We generalize it to the noncommutative theory as
 \eqn\vecs{\eqalign{
 V(y, \theta,\thetabar)=&
 -\theta \sigma^\mu\thetabar A_\mu(y) + i \theta\theta
 \thetabar \bar \lambda(y) -i\thetabar\thetabar \theta^\alpha
 \left( \lambda_\alpha(y) + {1 \over 4} \epsilon_{\alpha\beta}
 C^{\beta \gamma} \sigma^\mu_{\gamma\gammadot} \{\bar
 \lambda^\gammadot,A_\mu\}\right) \cr
 &+ {1\over 2} \theta\theta \thetabar\thetabar\left[D(y)- i
 \partial_\mu A^\mu(y)\right].}}
For nonzero $C$ we parameterized the $\thetabar\thetabar \theta $
term differently than in the commutative theory. The reason for
this will become clear momentarily.

Because of the noncommutativity the powers of $V$ \vecs\ are
deformed from their standard expressions
 \eqn\powve{\eqalign{
 &V^2= \thetabar \thetabar \left[-{1\over 2} \theta\theta A_\mu
 A^\mu - {1\over 2} C^{\mu\nu} A_\mu A_\nu +{i\over
 2}\theta_\alpha C^{\alpha\beta} \sigma^\mu_{\beta \alphadot}
 [A_\mu, \bar\lambda^{\alphadot}]-
 {1\over 8} |C|^2 \bar \lambda \bar \lambda \right] \cr
 &V^3= 0.
 }}

The remaining infinitesimal gauge symmetry \gauai, which preserves
the gauge choice \vecs\ is
 \eqn\remgau{\eqalign{
 \Lambda(y,\theta) = &- \varphi(y) \cr
 \bar\Lambda(\bar y,\theta) = &-\varphi(\bar y) - {i\over
 2} \thetabar\thetabar C^{\mu\nu} \{\partial_\mu \varphi,
 A_\nu\}\cr
 = & - \varphi( y) +2 i\theta\sigma^\mu\thetabar \partial_\mu
 \varphi (y) - \theta\theta\thetabar\thetabar \partial^2 \varphi(y)
 -{i\over 2} \thetabar\thetabar C^{\mu\nu} \{\partial_\mu \varphi,
 A_\nu\}.
 }}
Note the new term proportional to $C$ in $\bar \Lambda$ which is
needed in order to preserve \vecs. It is straightforward to check
that this transformation acts on the component fields as
 \eqn\gauic{\eqalign{
 &\delta A_\mu=-2\partial_\mu \varphi +i [\varphi, A_\mu]
 \cr
 &\delta \lambda_\alpha = i [\varphi,
 \lambda_\alpha]\cr
 &\delta \bar\lambda_\alphadot = i [\varphi, \bar
 \lambda_\alphadot]\cr
 &\delta D = i [\varphi, D]
 }}
i.e.\ as standard gauge transformations.

This discussion of the gauge symmetry explains why we
parameterized the coefficient of $\thetabar\thetabar\theta$ in $V$
as we did.  We wanted to ensure that the component fields
transform as in ordinary gauge theory.  This parametrization and
the special $\Lambda$ and $\bar \Lambda$ \remgau\ are similar in
spirit to the change of variables of  \SeibergVS\ from
noncommutative to commutative gauge fields in ordinary
noncommutative gauge theories.

We now turn to the field strengths. Since in the gauge \vecs\ $V$
has at least one $\thetabar$, the same is true for $D_\alpha V$.
Using these facts, it is easy to show that $e^{-V} D_\alpha
e^V=D_\alpha V + \half [D_\alpha V,V]$ (all the products are star
products), and therefore
 \eqn\Walpha{\eqalign{
 W_\alpha&=-{1\over 4} \bar D^2 e^{-V} D_\alpha e^V=-{1\over 4}
 \bar D^2 \left(D_\alpha V + \half [D_\alpha V,V]\right)\cr
 &=-i \lambda_\alpha(y) +\left[\delta_\alpha^\beta
 D (y)-i \sigma^{\mu\nu~ \beta}_\alpha \left( F_{\mu\nu}(y)
 + {i \over 2} C_{\mu\nu} \bar\lambda\bar
 \lambda(y)\right) \right]\theta_\beta + \theta\theta
 \sigma^\mu_{\alpha\alphadot} {\cal D}_\mu \bar
 \lambda^\alphadot(y)\cr
  &=W_\alpha(C=0) +\epsilon_{\alpha\gamma} C^{\gamma\beta}
  \theta_\beta \bar\lambda\bar \lambda(y) .
 }}
In this normalization the field strength and the covariant
derivative are
 \eqn\fiec{\eqalign{
 &F_{\mu\nu}=\partial_\mu A_\nu -\partial_\nu A_\mu +{i\over 2}
 [A_\mu,A_\nu]\cr
 &{\cal D}_\mu\lambda_\alpha = \partial_\mu \lambda_\alpha
 + {i\over 2} [A_\mu,\lambda_\alpha].}}
As we commented above, a more standard convention for gauge fields
is obtained with the substitution $V\to -2V$.

As a check, we note that under the infinitesimal gauge
transformation \remgau\
 \eqn\gaugeW{\delta W_\alpha(y,\theta)=i [\varphi(y),
 W_\alpha(y,\theta)].}

In order to compute $\bar W_\alphadot$ we use the fact that in the
gauge \vecs\ we have $e^V\bar D_\alphadot e^{-V}=-\bar D_\alphadot
V + \half [\bar D_\alphadot V,V] + \half V\bar D_\alphadot V V$,
where all the products are star products.  The last term vanishes
in the commutative theory, but it contributes for nonzero $C$. A
somewhat lengthy calculation leads to
 \eqn\Walphadot{\eqalign{
 \bar W_\alphadot&={1\over 4} D D e^V \bar D_\alphadot e^{-V}=
 -{1\over 4} D D \left(\bar D_\alphadot V -\half [\bar D_\alphadot
  V,V] - \half V\bar D_\alphadot V V \right)\cr
 &= \bar W_\alphadot(C=0) - \thetabar\thetabar \left[
 {C^{\mu\nu} \over 2} \{F_{\mu\nu},\bar \lambda_\alphadot\} +
 C^{\mu\nu} \{ A_\nu, {\cal D}_\mu \bar \lambda_\alphadot
 - {i\over 4}[A_\mu ,\bar \lambda_\alphadot]\}
 +{i\over 16}  |C|^2  \{\bar \lambda \bar
 \lambda, \bar \lambda_\alphadot\}\right].
 }}
As a check, we note that the infinitesimal gauge transformation
$\delta \bar W_\alphadot = -i[\bar \Lambda,\bar W_\alphadot]$ with
a star product with $\bar \Lambda$ of \remgau\ is
 \eqn\wbarg{\delta \bar W_\alphadot = i[\varphi(y),\bar
 W_\alphadot] +\thetabar\thetabar C^{\mu\nu}\left(2\{
 \partial_\nu\varphi, \partial_\mu \bar \lambda_\alphadot\}
 + {i \over 2}\left[\{\partial_\nu\varphi,A_\mu\},\bar
 \lambda_\alphadot\right]\right).}
It is implemented by the standard gauge transformations on the
component fields \gauic.

In order to find the Lagrangian we need
 \eqn\prelag{\eqalign{
 &\int d^2\theta ~\tr~ W W = \int d^2\theta~ \tr~ WW(C=0)
  -i C^{\mu\nu}\tr ~ F_{\mu\nu} \bar\lambda\bar\lambda +
  {|C|^2 \over 4} \tr ~ (\bar\lambda\bar\lambda)^2 \cr
 &\int d^2\thetabar ~  \tr ~\bar W \bar W= \int d^2\thetabar ~
 \tr ~\bar W \bar W(C=0) -i C^{\mu\nu}\tr ~ F_{\mu\nu}
  \bar\lambda\bar\lambda + {|C|^2 \over 4} \tr ~
  (\bar\lambda\bar\lambda)^2\cr
  &\qquad\qquad\qquad\qquad\qquad\qquad
   + {\rm total ~ derivative}.
 }}
The deformations of $\int d^2\theta ~\tr~ W W$ and of $\int
d^2\thetabar ~  \tr ~\bar W \bar W$ are equal up to total
derivatives.  Therefore, the deformation of the Lagrangian depends
only on ${\rm Im}~ \tau = {4\pi \over g^2}$.  More precisely, the
term multiplying ${\rm Re} ~\tau ={\theta \over 2\pi}$ remains a
total derivative.

Since the unbroken supersymmetry acts as a simple differential
operator $Q_\alpha={\partial \over \partial \theta^\alpha}$ (with
fixed $y$), it is easy to see how the various fields transform
 \eqn\Wmul{\eqalign{
 &\delta\lambda =i\epsilon D +\sigma^{\mu\nu } \epsilon
 \left( F_{\mu\nu} + {i \over 2} C_{\mu\nu}\bar\lambda\bar\lambda
 \right)\cr
 &\delta A_\mu=-i\bar \lambda \bar \sigma_\mu \epsilon \cr
 & \delta F_{\mu\nu} = i\epsilon\left(\sigma_\nu{\cal D}_\mu
 -\sigma_\mu{\cal D}_\nu \right) \bar\lambda\cr
 &\delta D=-\epsilon \sigma^\mu{\cal D}_\mu \bar \lambda\cr
 &\delta \bar \lambda =0.}}
The only effect of the deformation is the correction to $\delta
\lambda$.

\newsec{(Anti)Chiral Rings}

In this section we will explore some of the consequences of the
unbroken supersymmetry $Q$.  We will examine whether we can define
chiral and antichiral rings, and which of their properties in the
commutative theory survive in the noncommutative theory.

We start by considering the antichiral operators \chib\
$\bar\Phi(\bar y = y -2i\theta\sigma \thetabar,\thetabar)$.  Their
correlation functions are
 \eqn\corrc{\langle \bar\Phi_1(\bar y_1,\thetabar_1) \bar\Phi_2
 (\bar y_2,\thetabar_2) ... \bar\Phi_n(\bar y_n,\thetabar _n)
 \rangle = \bar F(\bar y_i = y_i -2i\theta_i\sigma
 \thetabar_i,\thetabar_i).}
Translation invariance shows that $\bar F$ depends only on $\bar
y_i - \bar y_j$.  If the vacuum is $Q$ invariant, we derive the
Ward identity
 \eqn\Qwi{\sum_i {\partial \over
 \partial \theta_i} \bar F=-2i\sigma^\mu \sum_i \thetabar_i
 {\partial \over \partial\bar y^\mu_i}\big|_{\thetabar_i} \bar
 F=0.}
This means that $\bar F(\bar y_i, \thetabar_i=0)$ is a constant
independent of $\bar y_i$. Therefore, the correlation functions of
first components of anti-chiral operators are independent of
separations.  Following the standard argument based on cluster
decomposition, the correlation function \corrc\ factorizes
 \eqn\corrcf{\langle \bar\Phi_1(\bar y_1,\thetabar_1=0) \bar\Phi_2
 (\bar y_2,\thetabar_2=0)  ... \bar\Phi_n(\bar y_n,\thetabar _n=0)
 \rangle =\prod_{i=1}^n\langle \bar\Phi_i(\bar y_i,\thetabar_i=0)
 \rangle ,}
where each expectation value in the right hand side is independent
of $\bar y_i$.  These correlation functions do not exhibit the
full complexity of quantum field theory involving short distance
singularities etc..  In other words, it is easy to multiply such
operators and they form a ring.

The set of antichiral operators which are $D$ derivatives of other
operators can be set to zero in the ring.  To see that, consider
the correlation function of the antichiral operator $\bar \Phi_1=
D_\alpha \Psi ( \bar y_1,\theta_1,\thetabar_1) $ for some $\alpha$
and $\Psi$. In this case the correlation function \corrc\ can be
written as
 \eqn\wis{\eqalign{
 \bar F(\bar y_1,...,\bar y_n,\thetabar_1,...,
 \thetabar_n)&=\left({\partial \over \partial \theta_1^\alpha}
 + 2 i \sigma^\mu _{\alpha\alphadot} \thetabar_1^\alphadot{\partial
 \over \partial y_1^\mu} \right)G( \theta_1,
 \bar y_1,...,\bar y_n, \thetabar_1,..., \thetabar_n)\cr
 &= 2 i \sigma^\mu _{\alpha\alphadot} \sum_{i=1}^n
 \thetabar^\alphadot_i{\partial \over \partial {\bar y}^\mu_i}
 \big |_{\theta_i,\thetabar_i} G( \theta_1,\bar y_1,...,
 \bar y_n, \thetabar_1,..., \thetabar_n),}}
where $G= \langle \Psi(\bar y_1,\theta_1, \thetabar_1)
\bar\Phi_2(\bar y_2,\thetabar_2) ... \bar\Phi_n(\bar y_n,\thetabar
_n) \rangle $ can have explicit $\theta_1$ dependence.  In the
last step in \wis\ we used the $\N={1\over 2}$ supersymmetry ($Q$
supersymmetry) Ward identity \Qwi.  It follows from \wis\ that for
such $\bar \Phi_1$ the correlation function $\bar F (\bar
y_1,...,\bar y_n,\thetabar_1=0,..., \thetabar_n=0)=0$, and
therefore we can mod out the set of operators $\{\bar \Phi_i(\bar
y,\thetabar=0)\}$ by those operators which are of the form $\bar
\Phi = D_\alpha \Psi$.

The antichiral operators form a ring with such an ideal in the
commutative theory.  Surprisingly, this is also true in the
noncommutative theory.  This is closely related to our discussion
above explaining that the contribution of the anti-superpotential
$\int d^2\thetabar ~\bar W$ to the action is not deformed by $C$.
However, since $\int d^2\thetabar ~\tr~\bar W_\alphadot \bar
W^\alphadot$ is deformed by $C$, the ring is not exactly the same
as in the commutative theory.

It is instructive to repeat this discussion from a somewhat
different point of view.  Instead of defining antichiral operators
as we did, we can define antichiral operators $\bar \CO$ by
$[Q,\bar \CO\}=0 $.  Since $Q$ is conjugate to $D$ this definition
is conjugate to the one we used above.  Now it is easy to prove,
using the fact that $Q$ is a symmetry of the theory, that if the
vacuum is invariant under $Q$, the correlation function
 \eqn\idec{\langle [Q,M\} \bar \CO_2...\bar\CO_n\rangle = \pm
 \langle \bar\CO_1 [Q,\bar \CO_2\}...\bar\CO_n\rangle\pm...\pm
 \langle \bar\CO_1 \bar \CO_2...[Q,\bar\CO_n\}\rangle =0}
(the signs $\pm$ depend on whether $M$ and $\CO_i$ are bosonic or
fermionic), and therefore the antichiral operators of the form
$[Q,M\}$ can be set to zero in the correlation functions.
Therefore, we can identify $\CO \sim \CO + [Q,M\}$ for every $M$.
Furthermore, even though $\bar Q$ of \QQbar\ is not a symmetry,
since \QDrelations\ $\{\bar Q_\alphadot, Q_\alpha\} =
2i\sigma^\mu_{\alpha\alphadot}{\partial \over
\partial y^\mu}$, it is still true that for an antichiral operator
$\partial \bar \CO\sim [Q, [\bar Q, \bar \CO\}\}$. Using \idec,
the correlation functions of $\partial \bar \CO$ with other
antichiral operators vanish. Therefore, the correlation functions
of antichiral operators are independent of separations and they
factorize. We again conclude that the antichiral operators form a
ring.

These facts are not true for the chiral operators.  Consider the
expectation value of a product of chiral operators
 \eqn\corrc{\langle \Phi_1(y_1,\theta_1) \Phi_2(y_2,\theta_2) ...
 \Phi_n(y_n,\theta_n) \rangle =
 F(y_1,...,y_n,\theta_1,...,\theta_n).}
Translation invariance guarantees that $F$ depends only on
$y_i-y_j$.  If the vacuum preserves our unbroken supersymmetry
$Q$, the correlation function $F$ is annihilated by
$Q=\sum_i{\partial \over \partial \theta_i }$, and therefore is a
function of $\theta_i-\theta_j$.  In the commutative theory we can
also use invariance under $\bar Q =\sum_i \left(-{\partial
\over\partial \thetabar_i} +2i\theta_i\sigma ^\mu{\partial\over
\partial y^\mu_i}\right)$ to show that $F(y_1,...,y_n, \theta_1=0,
...,\theta_n=0)$ is independent of $y_i$.  But since $\bar Q$ is
not a symmetry of the noncommutative theory we cannot argue that
in our theory. Therefore, the correlation functions of first
component of chiral superfields depend on separations, and these
operators no longer form a ring.  Similarly, since $\bar Q$ is not
a symmetry, we cannot identify chiral operators which differ by
$\bar D$ of another operator.

\newsec{Instantons}

In this section we study instantons and anti-instantons and their
fermion zero modes.  The deformation of the action \prelag\ does
not affect the purely bosonic terms.  Therefore, the instanton
equation $F^+=0$ with nonzero $F^-$, and the anti-instanton
equation $F^-=0$ with nonzero $F^+$ are not modified.

In the commutative theory the instantons or the anti-instantons
break half of the sueprsymmetry.  The instantons with $F^+=0$
break the $\bar Q$ supersymmetries and the anti-instantons with
$F^-=0$ break the $Q$ supersymmetries.  These broken charges lead
to some fermion zero modes. In addition to these, there are other
fermion zero modes whose existence follows from the index theorem.

Our deformation of superspace breaks supersymmetry to $\N={1\over
2}$ supersymmetry -- only $Q$ is a symmetry but $\bar Q$ is
broken. Unlike the breaking in an instanton background, which is
spontaneous breaking, this breaking is explicit.  Let us examine
the interplay between these two effects.  The anti-instantons
spontaneously break the supersymmetry which is preserved by the
deformation $Q$. Therefore, $Q$ should lead to fermion zero modes.
The instantons, on the other hand, break a symmetry which is
already broken by the deformation $\bar Q$. Therefore, this
symmetry does not lead to zero modes.  The unbroken supersymmetry
$Q$ pairs fermion and boson nonzero modes.

Let us see how these general considerations are compatible with
the index theorem. The fermion equations of motion derived from
the action \prelag\ are
 \eqn\fereq{\eqalign{
 &\sigma^\mu {\cal D}_\mu \bar \lambda =0 \cr
 &\bar\sigma^\mu {\cal D}_\mu  \lambda =-C^{\mu\nu}F_{\mu\nu}^+
 \bar \lambda -i {|C|^2 \over 2} (\bar\lambda\bar\lambda)
 \bar\lambda.}}
An anti-instanton ($F^-=0$) has $\lambda$ zero modes, and $\bar
\lambda$ vanishes.  Therefore the right hand side of the second
equation is zero and the equations are exactly as in the
commutative theory.  The situation with instantons ($F^+=0$) is
different. For zero $C$ there are $\bar \lambda$ zero modes but no
$\lambda$ zero modes. When $C$ is nonzero this remains true in
accordance with the index theorem. However, at order $\bar
\lambda^3$ we find a source in the $\lambda$ equation and
$\lambda$ cannot remain zero. Therefore, the zero modes are not
lifted but it appears to be difficult to introduce collective
coordinates for them. This is consistent with the fact that they
are not associated with a broken symmetry.

\newsec{Relation to Graviphoton Background}

In this section we show how our noncommutative superspace arises
in string theory in background with constant graviphoton field
strength.  Our discussion is motivated by the work of Ooguri and
Vafa \OoguriQP, and is similar to it.  But we differ from
\OoguriQP\ in some crucial details.

We start with the holomorphic part of the heterotic strings using
Berkovits' formalism (for a nice review see \BerkovitsBF).  The
relevant part of the worldsheet Lagrangian is
 \eqn\wslag{\CL_h= {1\over \alpha'}\left({1\over 2} \tilde
 \partial x^\mu \partial x_\mu + p_\alpha \tilde \partial
 \theta^\alpha + \bar p_\alphadot
 \tilde \partial\thetabar^\alphadot\right), }
where we ignore the worldsheet fields $\rho$ and $\zeta$.  Since
we use a bar to denote the space time chirality, we use a tilde to
denote the worldsheet chirality.  Therefore when the worldsheet
has Euclidean signature it is parametrized by $z$ and $\tilde z$
which are complex conjugate of each other.  For a Lorentzian
signature target space $p_\alpha$ is the hermitian conjugate of
$-\bar p_\alphadot$.  For a Euclidean target space they are
independent fields.  $p$ and $\bar p$ are canonically conjugate to
$\theta$ and $\thetabar$; they are the worldsheet versions of
$-{\partial \over\partial \theta} \big|_x$ and $-{\partial \over
\partial \thetabar} \big|_x$.  The reason $x$ is held fixed in
these derivatives is that $x$ appears as another independent field
in \wslag.  We also define $d_\alpha$, $\bar d_\alphadot$,
$q_\alpha$ and $\bar q_\alphadot$, which are the worldsheet
versions of $D_\alpha$, $\bar D_\alphadot $, $Q_\alpha$ and $\bar
Q_\alphadot $
 \eqn\dqdef{\eqalign{
 d_\alpha &= -p_\alpha + i \sigma^\mu_{\alpha\alphadot}
  \thetabar^\alphadot \partial x_\mu -\thetabar\thetabar
 \partial\theta_\alpha +{1\over 2} \theta_\alpha \partial
 (\thetabar\thetabar)\cr
  \bar d_\alphadot &= \bar p_\alphadot -i \theta^\alpha
  \sigma^\mu_{\alpha\alphadot}\partial x_\mu - \theta\theta
 \partial\thetabar_\alphadot +{1\over 2} \thetabar_\alphadot
  \partial (\theta\theta)\cr
 q_\alpha  &=-p_\alpha -i \sigma^\mu_{\alpha\alphadot}
 \thetabar^\alphadot \partial x_\mu +{1\over 2}
 \thetabar\thetabar \partial\theta_\alpha -{3\over 2}\partial
 (\theta_\alpha \thetabar\thetabar)\cr
 \bar q_\alphadot &= \bar p_\alphadot +i\theta^\alpha
 \sigma^\mu_{\alpha\alphadot} \partial x_\mu +{1\over 2}
 \theta\theta \partial\thetabar_ \alphadot -{3\over 2}
 \partial ( \thetabar_\alphadot\theta\theta)  .}}
Our definitions of $q$ and $\bar q$ differ from the inetgrand of
the supercharges in \BerkovitsBF\ by total derivatives which do
not affect the charges, but are important for our purpose.

Since we are interested in working in terms of $y^\mu=x^\mu+i
\theta^\alpha\sigma^\mu_{\alpha\alphadot}\thetabar^\alphadot $,
and the derivatives with fixed $y$, $Q_\alpha={\partial \over
\partial \theta}\big|_y$ and $\bar D_\alphadot=- {\partial \over
\partial \thetabar}\big|_y$, we change variables to $y$,
$q_\alpha$ and $\bar d_\alphadot$.  A simple calculation leads to
 \eqn\wslaga{\CL_h={1\over \alpha'}\left({1\over 2} \tilde
 \partial y^\mu \partial y_\mu - q_\alpha \tilde \partial
 \theta^\alpha + \bar d_\alphadot \tilde \partial
 \thetabar^\alphadot + {\rm total ~derivative}\right).}
In deriving this expression the total derivatives in $q$ in
\dqdef\ is important.

In the type II theory \wslag\ is replaced with
 \eqn\wslagII{ \CL_{II} = {1\over \alpha'}\left({1\over 2} \tilde
 \partial x^\mu \partial x_\mu + p_\alpha \tilde \partial
 \theta^\alpha + \bar p_\alphadot \tilde
 \partial\thetabar^\alphadot + \tilde p_\alpha \partial
 \tilde\theta^\alpha + \tilde {\bar p}_\alphadot
 \partial\tilde\thetabar^\alphadot\right),}
where again the tilde denotes the worldsheet chirality. Changing
variables to
 \eqn\chav{\eqalign{
 y^\mu &=x^\mu+i \theta^\alpha\sigma^\mu_{\alpha\alphadot}
 \thetabar^\alphadot  +i \tilde \theta^\alpha\sigma^
 \mu_{\alpha\alphadot}\tilde\thetabar^\alphadot\cr
 \bar d_\alphadot &= \bar p_\alphadot -i \theta^\alpha
 \sigma^\mu_{\alpha\alphadot}\partial x_\mu - \theta\theta
 \partial\thetabar_\alphadot +{1\over 2} \thetabar_\alphadot
 \partial (\theta\theta)\cr
 q_\alpha  &=-p_\alpha -i \sigma^\mu_{\alpha\alphadot}
 \thetabar^\alphadot \partial x_\mu +{1\over 2}
 \thetabar\thetabar \partial\theta_\alpha -{3\over 2}\partial
 (\theta_\alpha \thetabar\thetabar) \cr
 \tilde{\bar d}_\alphadot &= \tilde{ \bar p}_\alphadot -i \tilde
 \theta^\alpha \sigma^\mu_{\alpha\alphadot}\tilde \partial x_\mu
 - \tilde\theta\tilde\theta \tilde \partial\tilde
 \thetabar_\alphadot +{1\over 2} \tilde\thetabar_\alphadot
  \tilde \partial (\tilde\theta\tilde\theta)\cr
 \tilde q_\alpha  &=-\tilde p_\alpha-i\sigma^\mu_{\alpha
 \alphadot}\tilde \thetabar^\alphadot \tilde \partial x_\mu
 +{1\over 2} \tilde\thetabar\tilde \thetabar \tilde \partial
 \tilde \theta_\alpha -{3\over 2}\tilde \partial (\tilde
 \theta_\alpha \tilde \thetabar\tilde \thetabar)}}
we derive
 \eqn\wslagIIa{{\cal L}_{II}={1\over \alpha'}\left({1\over 2} \tilde
 \partial y^\mu \partial y_\mu -q_\alpha \tilde \partial
 \theta^\alpha
 + \bar d_\alphadot \tilde \partial\thetabar^\alphadot - \tilde
 q_\alpha \partial \tilde \theta^\alpha + \tilde{\bar d}_\alphadot
 \partial\tilde \thetabar^\alphadot + {\rm total
 ~derivative}\right).
 }

If the worldsheet ends on a D-brane, the boundary conditions are
easily found by imposing that there is no surface term in the
equations of motion.  For a boundary at $z=\tilde z$, we can use
the boundary conditions $\theta(z=\tilde z)=\tilde \theta(z=\tilde
z)$, $q(z=\tilde z)=\tilde q(z=\tilde z)$, etc.  Then the
solutions of the equations of motion are such that
$\theta(z)=\tilde \theta(\tilde z)$, $q(z)=\tilde q(\tilde z)$,
etc.; i.e.\ the fields extend to holomorphic fields beyond the
boundary.  The boundary breaks half the supersymmetries preserving
only $\oint q dz + \oint \tilde q d\tilde z$ and $\oint \bar q dz
+ \oint \tilde {\bar q} d\tilde z$.

Motivated by \OoguriQP\ we consider the system in a background of
constant graviphoton field strength $F_{\alpha\beta}$. We take the
field strength $F$ to be selfdual ($F_{\alphadot\betadot}=0$)
because such a background is a solution of the spacetime equations
of motion without back reaction of the metric.  This background is
represented in the worldsheet Lagrangian by adding to \wslagII\ or
\wslagIIa\ the term
 \eqn\gravip{F^{\alpha\beta} q_\alpha\tilde q_\beta .}
The form \wslagIIa\ is particularly convenient because it makes it
manifest that the worldsheet theory remains free in this
background.  It is important that the coordinates $y$ are free and
independent of the background, while the original standard
spacetime coordinates $x$ couple to $F$.

Ignoring the trivial fields we are led to consider the
Lagrangian
  \eqn\wslagIIai{{1\over \alpha'}\left( -q_\alpha \tilde \partial
  \theta^\alpha  - \tilde q_\alpha \partial \tilde \theta^\alpha +
  \alpha'F^{\alpha\beta} q_\alpha\tilde q_\beta \right). }
The fields $q$ and $\tilde q$ can easily be integrated out using
their equations of motion
 \eqn\qtqeom{\eqalign{
 & \tilde \partial \theta^\alpha = \alpha' F^{\alpha\beta} \tilde
 q_\beta\cr
 &\partial\tilde \theta^\alpha = -\alpha' F^{\alpha\beta}
 q_\beta\cr}}
leading to
  \eqn\wslagIIaii{{\cal L}_{eff}= \left({1\over \alpha'^2 F}
  \right)_{\alpha\beta}  \partial\tilde \theta^\alpha \tilde
  \partial   \theta^\beta . }
When we consider a system with a boundary (along $z = \tilde z$
for a Euclidean worldsheet) we need to find the appropriate
boundary conditions.  These are determined from the surface term
in the equations of motion
 \eqn\sureom{\left({1\over F} \right)_{\alpha\beta}\left( \tilde
 \partial \theta^\alpha \delta \tilde \theta^\beta+  \partial
 \tilde \theta^\alpha \delta \theta^\beta\right)
 \big|_{z=\tilde z} =0.}
We impose
 \eqn\bouncondi{\theta^a (z=\tilde z) =\tilde \theta^a (z=
 \tilde z), \qquad \partial \tilde \theta ^a(z=\tilde z)
 = -\tilde \partial \theta^a(z=\tilde z).}
The first condition states that the superspace has half the number
of $\theta$s.  The second condition guarantees, using \qtqeom\
that $q_\alpha(z=\tilde z)=\tilde q_{\alpha}(z=\tilde z)$.

The various propagators are found by imposing the proper
singularity at coincident points, Fermi statistics and the
boundary conditions \bouncondi.  We find
 \eqn\props{\eqalign{
 &\langle \theta^\alpha(z,\tilde z) \theta^\beta(w,\tilde
 w)\rangle = {\alpha'^2F^{\alpha\beta} \over 2\pi i}
 \log{\tilde z-w\over z-\tilde w}\cr
 &\langle \tilde\theta^\alpha(z,\tilde z) \tilde\theta^\beta
 (w,\tilde w)\rangle = {\alpha'^2 F^{\alpha\beta} \over 2\pi i}
 \log{\tilde z-w\over z-\tilde w}\cr
  &\langle \theta^\alpha(z,\tilde z) \tilde\theta^\beta
 (w,\tilde w)\rangle = {\alpha'^2 F^{\alpha\beta}\over 2\pi i}
 \log{(z-w)(\tilde z-\tilde w)\over (z-\tilde w)^2}.
 }}
The branch cuts of the logarithms is outside the worldsheet.
Therefore for two points on the boundary $z=\tilde z=\tau$ and
$w=\tilde w= \tau'$
 \eqn\bounpro{\langle \theta^\alpha(\tau)\theta^\beta(\tau')
 \rangle = {\alpha'^2 F^{\alpha\beta} \over 2} {\rm sign}
 (\tau-\tau').}
Using standard arguments about open string coupling, this leads to
 \eqn\noncog{\{\theta^\alpha,\theta^\beta\}=\alpha'^2
 F^{\alpha\beta}=C^{\alpha\beta} \not=0;}
i.e.\ to a deformation of the anticommutator of the $\theta$s.  It
is important that since the coordinates $\bar \theta$ and $y$ were
not affected by the background coupling \gravip, they remain
commuting. In particular we derive that $[y^\mu,y^\nu]=0$ and
therefore $[x^\mu,x^\nu]\not=0$, exactly as motivated earlier by
consistency.

We conclude that the graviphoton background leads to exactly the
same deformation of superspace which we considered above.

Since the equation of motion of $\theta$ states that $q$ is
holomorphic, as in the case without the background, it extends to
a holomorphic field $\tilde q (\tilde z)= q(z)$, and therefore
$\oint q dz + \oint \tilde q d\tilde z$ is a conserved charge.
However, even though $\partial\tilde \theta \sim q$ is holomorphic
and $\tilde \partial \theta \sim \tilde q$ is antiholomorphic,
$\theta$ and $\tilde \theta$ are no longer holomorphic and do not
extend holomorphically through the boundary (see the propagators
\props). Therefore, the supersymmetries $\oint \bar q dz + \oint
\tilde{\bar q}d\tilde z $ are broken by the deformation.

\bigskip
\centerline{\bf Acknowledgements}

It is a pleasure to thank  N. Berkovits, F. Cachazo, S. Cherkis,
M. Douglas, R. Gopakumar, A. Hashimoto, J. Maldacena, N. Nekrasov,
D. Shih and E. Witten for helpful discussions. This work was
supported in part by DOE grant \#DE-FG02-90ER40542 and NSF grant
PHY-0070928 to IAS.

\listrefs

\end